\documentclass[12pt,preprint]{aastex}

\shorttitle{Spectral Variation in Emission and Absorption in Vela X-1 by Orbital Phase}
\shortauthors{Goldstein et al.}

\begin{document}

\title{Variation in Emission and Absorption Lines and Continuum Flux by 
Orbital Phase in Vela X-1}

\author{Greg Goldstein\altaffilmark{1}}
\affil{Astronomy Department, University of Western Sydney
    Sydney, Australia}
\email{greggold@bluewin.ch}

\author{David P. Huenemoerder}
\affil{MIT Center for Space Research, 70 Vassar St., Cambridge MA 02139}
\email{dph@space.mit.edu}

\and

\author{David Blank}
\affil{School of Engineering and Industrial Design, University of Western Sydney
    Sydney, Australia}
\email{rlagn@yahoo.com}

\altaffiltext{1}{doctoral student, Centre for Astronomy, James Cook University}

\begin{abstract}

High resolution spectral studies were undertaken at orbital phases ($\phi$) 0, 0.25 and
0.5 on the high-mass X-ray binary (HMXB) Vela X-1 using archival Chandra data.
We present (a) the first detailed analysis of the multiple strong narrow emission lines present
in $\phi$ 0.5  (b) an analysis of the absorption of the continuum in  $\phi$ 0.5, and (c) the first
detection of narrow emission and absorption lines in $\phi$ 0.25. Multiple fluorescent and H-and
He-like emission lines in the band 1.6 - 20 Angstrom (\AA) in eclipse are partially obscured at
$\phi$ 0.25 by the X-ray continuum. The $\phi$ 0.25 spectrum displays 3 triplets, 2 with a
blue-shifted resonance (r) line in absorption and the intercombination (i) and forbidden (f) lines
in emission, and shows in absorption other blue-shifted lines seen in emission in eclipse.
At $\phi$ 0.5 the soft X-ray continuum diminishes revealing an "eclipse-like" spectrum, however
line flux values are around 13-fold those in eclipse. We conclude the narrow emission lines in Vela
X-1 become apparent when the continuum is blocked from line of sight, either by eclipse or by
scattering and/or absorption from a wake or cloud. The H-and He-like lines arise in warm
photoionised regions in the stellar wind, while the fluorescent lines (including a Ni K$\alpha$
line) are produced in cooler clumps of gas outside these regions. Absorption of the 5-13 $\mbox{\AA}
$ continuum at $\phi$ 0.5 may be caused by an accretion wake comprised of dense stagnant
photoionized plasma inside a Stromgren zone. Multiple fluorescent emission lines may be a common
feature of the supergiant category of HMXBs.

\end{abstract}


\section{Introduction}

High-mass X-ray binaries (HMXBs) including Vela X-1 have served to illuminate
the evolution of giant binary stars, to probe their radiation-driven winds, and
to study accretion processes and the fundamental parameters of compact
objects. The 130 known HMXBs fall into 2 categories, Be/X-ray and supergiant
systems \citep{liu00}. Vela X-1 is one of 11 supergiant systems (which
include GX 301-2 and 4U 1700-37), while the vast majority (around 80\%) of
HMXBs are Be/X-ray systems detected as transients \citep{kap01,whi95}.
Supergiant systems have a relatively short life of 10,000 years before the
secondary star explodes in a supernova leaving a neutron star binary or 2
single neutron stars \citep{kap01}. Once a massive main-sequence star evolves
to a supergiant the Roche lobe may become filled and accretion flow becomes
sufficient to power a strong X-ray source.

Vela X-1 is an eclipsing system at an estimated distance of 1.9 kpc
\citep{sad85} containing an B0.5Ib supergiant (mass 23 $M_\odot$ and radius 34
$R_\odot$) and a wind-fed pulsar with a period of 283 s \citep{nag89}. The
neutron star (NS) orbits in a period of 8.96 days very close to the
non-synchronously rotating surface of the supergiant with a semi-major axis
53.6 $R_\odot$\citep{bar01}, so it is fully embedded in the acceleration zone
of the stellar wind.  The NS mass of 1.9 $M_\odot$ is the highest known
\citep{bar01,van95} and the system has been studied at infrared \citep{hut02},
UV \citep{sad85, vanl01}, gamma \citep{rau94} and at optical \citep{van95,
kap94} wavelengths. It has not been detected at radio frequency (detection
limit at mean cm radio flux density (mJy) $<0.2$ \citep{fen00}.

Orbital modulation of the spectra have been noted in the UV, X-ray, and optical
wavelengths. 
\citet{sad85} noted absorption of \ion{Al}{3} and \ion{Fe}{3} in the UV,
limited to the orbital phase ($\phi$) 0.5 and higher, and postulated the presence of a
trailing wake. \citet{bes75} reported an absorption component in the H$\alpha$
spectrum around $\phi$ 0.6.  \citet{kap94} postulated that an absorption
component in the optical spectrum resulted from a photoionization wake that
trails the X-ray source, rather than the accretion wake
which does not sufficiently obscure the supergiant. \citet{smi01} studied an
asymmetric dip in the UV lightcurve at $\phi$ 0.46-0.7 and concluded
spectral ratios (predip/dip) contain information about the mechanism
responsible for the variability. Absorption of soft X-rays during the $\phi$
0.5 and higher was first reported in Tenma data \citep{nag89}. 
\citet{fel96} showed that a trailing photoionization wake can explain the
asymmetry in the X-ray band during the eclipse ingress and egress.

\citet{sak99} using ASCA data demonstrated that the spectrum in eclipse was
comprised of recombination lines and radiative combination continua
(RRC) produced by photoionization in the stellar wind; they suggested
the fluorescent K-shell lines from near-neutral atoms indicated that
the X-ray -irradiated portion of the wind in Vela X-1 consists of cool
dense clumps in a hotter, more ionized gas.

\citet{shu02a} presented the first high-resolution spectrum during eclipse, and
using RRC (including \ion{Ne}{10} RRC) deduced a
temperature of 1.2 $\times 10^5$ K for the plasma. The spectrum showed clear
evidence of photoionization processes, and the resonance (r) lines in the
He-like triplets were of roughly equal strength to the forbidden (f) 
lines, consistent with resonant scattering.

We unexpectedly found that in $\phi$ 0.5 the  emission lines were stronger than in eclipse by an
order of magnitude, associated with the near disappearance of the soft X-ray continuum.  This
result, together with the unusual finding of multiple fluorescent emission lines, led us to further
investigate emission line production in HMXBs by a study of Vela X-1 comparing eclipse, $\phi$ 0.25
and $\phi$ 0.5 using high-resolution Chandra datasets. We outline the data reduction proceedure in
section 2, give our results in section 3 and present the discussion in section 4.

\section{Observations}

An analysis of Vela X-1 in eclipse (Chandra Dataset ObsId 102) has previously
been reported \citep{shu02a}. Three newer datasets obtained during one binary
orbit (Figure 1) in 2001 with the Chandra High Energy Transmission Grating
Spectrometer (HETGS) representing the orbital phases eclipse ($\phi$ 0),
$\phi$ 0.25 and $\phi$ 0.5 were available from the public archive (ObsIDs 1926, 1928,
1927, respectively (see Table~\ref{tbl-1}). These datasets have not been previously
reported, except for a specialized study of pulse-phased resolved emission of the Fe K$\alpha$ line
\citep{pau02}, and a summary report in a review of stellar winds in 5 HMXBs
\citep{sak03}.
The telescope operated in  "faint" data mode, with a timed readout
(frame time) of 1.74 seconds. The HETGS employs two sets of  transmission gratings: the Medium 
Energy Grating (MEG)
with a range of 2.5-31 Angstroms (\AA) (0.4-5.0 keV) and the High Energy Grating
(HEG) with a range of 1.2-15 $\mbox{\AA} $    (0.8-10.0 keV).
 The resolution  of the  first-order HEG
spectrum is $\bigtriangleup\lambda$ = 0.012  $\mbox{\AA} $, and MEG spectrum
$\bigtriangleup\lambda$ = 0.023 \AA. The dispersed spectra were detected with the ACIS-S
linear array of 6 CCD detectors. The HETGS/ACIS-S combination provides an undispersed
(zeroth order) image and dispersed spectra from the gratings. The zeroth-order image of
the non-eclipse datasets were affected by photon pileup and were not used in spectral
analysis.

A complete re-processing was undertaken using Ciao software
(version 2.3 with Caldb version 2.22) (including the tools acis\_process\_events,
tgdetect etc) (Chandra X-ray Center Data
analysis\footnote{http://asc.harvard.edu/ciao/}) to produce grating
ancillary response files (ARFs) and redistribution matrix files (RMFs). For
spectral response the files acisheg1D1999-07-22rmfN0004.fits and
acismeg1D1999-07-22rmfN0004.fits were used for the HEG
and MEG respectively. 

Measurement of emission lines in photoionized spectra involves consideration
of the contribution to the lines by a continuum, and using Isis (version
1.0.50) lines were fitted with a gaussian plus a polynomial to fit the narrow
regions adjacent to the line.

Initial inspection of the $\phi$ 0.25 spectrum showed a jagged appearance with
few apparent narrow features apart from the Fe K$\alpha$ line. For all datasets
a fast fourier transform routine in Isis was used to smooth the spectral
curve, which was viewed "in counts" to assess statistical quality and find
instrumental features, "in flux density" to compare HEG and MEG spectra, and in "delta-chi" to see the relative
significance of deviations. This procedure identified a number of previously
unreported emission and absorption features in the $\phi$ 0.25 spectrum,
including 3 He-like triplets. "Features" that do not appear both in HEG and
MEG spectra are likely noise, or are more prominent in the HEG spectrum due to its higher
resolution (Figures 2,3). Broad systematic flux differences are due to residual calibration
errors. For $\phi$ 0.25 spectra,  the statistical uncertainties are about 6\% and 11\% for
MEG and HEG, respectively, near 7.5 \AA, and rise to about 18\% and 32\% near 11.5 \AA.

Pileup in a Charge-Coupled Device (CCD) occurs when there is
coincidence of two or more photons per CCD frame-time within an event detection
cell. Two or more photons interact in a detection cell of the detector
resulting in a pulse height that is roughly the sum of the pulse heights
of the individual photon events.This causes a lower event detection
rate and distortion of the observed spectrum towards higher energies \citep{dav01}.
The zero-order images of ObsID 1928 and 1927 showed visual evidence of pile-up.
We checked for pileup in Vela X-1 (ObsId1928) first dispersion order very carefully, as its flux
and mean MEG count rates are comparable to another X-ray binary 4U 1543-624 (ObsId 702) which is
known to have mild pileup in the first order.
The important metric in determining pileup is the counts/frame/cell, where a cell
is a 3 pixel bin along the spectral dispersion axis  (Dewey D
\footnote{http://space.mit.edu/HETG/technotes/pileup/pileup\_9912.html})

Vela X-1 has 0.0345 and 4U 1543-624 has 0.045 counts/frame/cell, that is 4U 1543-624 has
counts/frame/cell 30\% higher than Vela X-1. Dewey argues the value of this metric indicates the magnitude of pileup effects: for
values of 0.01 and less pileup effects are negligible; for values of 0.01 to 0.1 pileup
may produce changes in event rates less than 10\%; for values above 0.1 pileup can be
severe and the data quality dubious.

The lack of pileup in Vela X-1 is demonstrated  by a careful comparison of the HEG and MEG
datasets in the range 1.7-15.2 \AA. There is close agreement in
the first-order spectra. In a graded series of 5 spectra with varying degrees of pileup,
all spectra with pileup (including 4U 1543-624, the mildest case of the 5) show lack of close
agreement between the MEG and HEG (J Davis
\footnote{http://cxc.harvard.edu/ccw/proceedings/02\_proc/presentations/j\_davis/pileup.html}).
The MEG has a larger effective area than HEG, and in the event of pileup
the MEG will be piled (or more piled) than HEG, and the MEG spectrum saturates and does not
follow (shows lower flux) than the HEG.

The assessment of absorption features in the $\phi$ 0.25 spectrum found
photoelectric neutral K absorption edges from S, Si and Mg in cold gas. An
optical depth (tau) of the neutral edge was assessed:

    $\tau$ = $\ln(f_{high}/f_{low})$;

where $"f_{high}"$ is the X-ray flux on top of the edge at the high
wavelength side and $"f_{low}"$ is the X-ray flux at the bottom of the
edge at the low wavelength side \citep{shu02b}. This optical depth is
directly proportional to neutral absorption column densities and
assesses the total amount of photo-electric absorption, which includes
absorption in the vicinity of the source as well as the interstellar
medium in the line of sight.

\citet{smi01} in studying the dip in the UV lightcurve at $\phi$
0.46-0.7 compared spectra predip to dip to obtain a "ratioed spectrum":

$F_{obs}(\lambda) = 1-f_{min}(\lambda)/f_{max}(\lambda)$

Between 5 and 13 \AA , the strong X-ray continuum noted in $\phi$ 0.25
was markedly attenuated in $\phi$ 0.5. We consider a simple hypothesis in which
an absorber (accretion wake associated with Bondi-Hoyle accretion)
causes absorption at $\phi$ 0.5, but is not in the line-of sight at $\phi$ 0.25
(Figure 4). We estimated the "ratioed spectrum," and also modeled an optical
depth of the absorption using:

$\tau = \-ln(f_{phase0.25} /f_{phase0.5})$.

\section{Results}

The 3 datasets were observed during 1 orbit of the companion by the NS (Figure
1) and no flaring behaviour occurred. The lightcurves from the 3 observations
are compared in Figure 5, in which pulsed emission (period 283 s) may be
discerned.

The 0.5-10 keV spectra were fit with a model of 2 absorbed powerlaws, a
scattered and direct
component \citep{sak99}. The photon index was fixed to 1.7 in both powerlaws in
eclipse,
with the second powerlaw index allowed to vary in the other phases (see
Table~\ref{tbl-1}, Figure 6). The total flux in $\phi$ 0.5 is reduced compared
to $\phi$ 0.25 (Table~\ref{tbl-1}, Figure 2).

Details of selected spectral lines and RRC are set out in
Tables~\ref{tbl-2},~\ref{tbl-3},\&~\ref{tbl-4}. A rich mosaic of 10 H-like, 12
He-like and 15 fluorescent lines (including the unusual Ni K$\alpha$
line), and 2 RRC were detected in the eclipse phase.  Our results in eclipse
phase are in general agreement with those reported by \citet{shu02a}, and were
not significantly different from ObsID 102.

Comparisons of 2 spectra (from $\phi$ 0.25 and 0.5) are presented in Figure
2 (wavelength range 1.5-14.3 \AA) and Figure 3 (1.6-2.0 \AA).

The $\phi$ 0.25 displays a mix of narrow features in emission and in absorption 
(Tables~\ref{tbl-2},~\ref{tbl-3}
; Figures 2,3), and 5 photo-electric absorption edges were detected
(Table~\ref{tbl-5}). The narrow emission and absorption lines in this phase
(including three He-like triplets) have not been previously reported. Two
triplets have blue-shifted r lines seen in absorption, while the
intercombination (i) and f lines are seen in emission (Figures 2,
Table~\ref{tbl-3}).

Comparing $\phi$ 0.5 to eclipse, the H- and He-like emission lines are 13 fold
stronger at $\phi$ 0.5 (mean value for comparisons on 11 lines where firm
confidence intervals established). The Fe K$\alpha$ is 21 fold
and Fe K$\beta$ 31 fold stronger. The Ca and Ar fluorescent lines seen in
eclipse are not found in $\phi$ 0.5, probably due to the stronger direct
continuum below 5 $\mbox{\AA} $ in $\phi$ 0.5. The Ni K$\alpha$ line and a Fe K
edge are shown in Figure 3. Vela X-1 is the only XRB with a Ni K$\alpha$ line
reported, however our study of GX 301-2 (in progress) has detected this line
(Table~\ref{tbl-2}).

In $\phi$ 0 and $\phi$ 0.5 the r and f lines of the He-like triplets have roughly
equal strength. However in $\phi$ 0.25  the triplet r line is a little
weaker than the f line (\ion{Ne}{9}) or is in absorption (\ion{Mg}{11},
\ion{Si}{13}) (Figure 2).

The He and H-like emission lines showed a red-shift at $\phi$ 0 (of around 300
km $s^{-1}$), and at $\phi$ 0.25 (of around 200 km $s^{-1}$). However lines
seen in absorption at $\phi$ 0.25 have a blue shift 
of around
300 km $s^{-1}$, consistent with the orbital velocity of the NS. At $\phi$ 0.5
the emission lines 
have a
blue-shift (of around 150 km $s^{-1}$).

The dramatic attenuation of the X-ray continuum at $\phi$ 0.5 compared to $\phi$
0.25 is modeled both as a spectral ratio ($\phi$ 0.25/$\phi$ 0.5) and as an
optical depth in Figure 7.  The optical depth increases steeply between 3 and 6 \AA, then gradually
declines from 6-13 \AA.

\section{Discussion}

These results provide insights into how narrow lines are produced in HMXBs, and
particularly into multiple fluorescent emission and orbital changes in
emission line strength. The main features to be discussed below are:

(a) changes with orbital phase in emission lines and the strong narrow emission lines present in
$\phi$ 0.5, the first detection of narrow emission and  absorption lines in $\phi$ 0.25, and an
analysis and measurement of the absorption of the X-ray continuum in $\phi$ 0.5.

(b) new aspects of the multiple fluorescent  emission in Vela X-1, including the identifications
of the Ni K$\alpha$ line in Vela X-1, and
also in GX301-2; the variation in line strength of Fe and Ni K$\alpha$ with orbital
phase, attributed (following Matt et al 1997) to an 
accreting system where the direct radiation is blocked, and only reflected
radiation is visible; 
and the finding that multiple fluorescent emission lines may be a common
feature of the 
supergiant category of HMXBs. 

(c) Doppler shifts of emission lines in $\phi$ 0.5 indicate a slowing of the
wind associated with the presence of the X-ray source and formation of the Stromgren zone and 
accretion wake. Two triplets in $\phi$ 0.25 show blue-shifted resonance lines seen in absorption
while the intercombination and forbidden lines are seen in emission, with the
blueshifted line in absorption having a different source to the lines seen in
emission.

 The X-ray luminosity of Vela X-1 is known to demonstrate considerable
variation on time intervals from hours to days \citep{kre99,van95}; there may be pulse-to-pulse
variations even though the pulse profile averaged over many pulses is quite stable.
Clearly our Chandra results (where flux change, and absorption change between phases are
carefully measured) relate to only the observed orbit.

\subsection{Photoionized Plasma }

In Vela X-1 spectral features indicating plasmas in photoionization equilibrium
(PIE) include (a) multiple fluorescent ions (b) narrow RRCs (\ion{Ne}{10} and
\ion{Ne}{9}) (c) relatively weak or apparent absence of iron L-shell emission
in presence of K-shell emission from lighter elements (S, Si, Mg, Ne and O)
\citep{lie01}, and (d) He-like triplet ratios that are consistent with 
photoionization and resonant scattering (involving photoexcitation) in a PI
plasma.  For \ion{Ne}{10} the Lyman series
(Ly$\alpha$:$\beta$:$\gamma$:$\delta$) have line ratios of 1.0:0.35:0.17:0.12
in eclipse and 
1.0:0.46:0.34:0.13 in $\phi$ 0.5. These flux ratios support the existence of
photoexcitation since 
the fluxes of the higher lines in the series are higher than expected from a
purely recombining 
or purely collisional plasma \citep{shu02a}.

PIE are characterised by  processes of photoionization (absorbed photon ejects
electron from ion) 
and recombination (electron recombines with ion and emits photon or excites
another electron). In 
PIE the ratio of the X-ray flux to the electron density is sufficiently high
that collisional 
ionization processes are negligible relative to photoionization processes
\citep{hee99}. Spectra 
of PIE consist of either recombination emission, or fluorescent emission, or a
composite of these 
\citep{lie99}. The dominance of fluorescence over recombination, or vice versa,
is related to the 
charge state distribution, with highly ionized plasmas being
recombination-dominated, while 
colder, less ionized plasma are fluorescence-dominated.

The He-like triplets have r and f line fluxes of roughly equal strength (except
in the $\phi$ 0.25, see section 4.4) (Table~\ref{tbl-3},Figure 2).
This is consistent with resonant scattering, defined as photoexcitation and
subsequent decay by
\citet{woj03}. The strong r line in eclipse in a photoionized plasma occurs
because the 
Ly$\alpha$ lines of H-like ions and the $n=2\rightarrow1$ r lines of He-like
triplets have large oscillator strengths (unlike the i and the f lines).
Line emission of lines with large
oscillator strengths is enhanced in the wind of an X-ray binary by
photoexcitation from radiation 
from the compact source. However outside of eclipse photons from the NS may be
scattered out of 
the line of sight to the observer, reducing the line flux of the r line of the
He-like triplet 
(and the Ly$\alpha$ lines), and if sufficient photons are scattered the r line
may be seen in 
absorption, as noted in $\phi$ 0.25 with \ion{Si}{13} and \ion{Mg}{11}
(Table~\ref{tbl-3}, Figure 2). In 
$\phi$ 0.5 the absorption of the continuum creates an "eclipse-like" spectrum
and triplets in $\phi$ 0 and 0.5 display similar ratios.

Narrow RRCs from \ion{Ne}{10} and \ion{Ne}{9} (at 9.10 $\mbox{\AA}$ and 10.37 \AA) were
detected in eclipse and $\phi$ 0.5 and indicate a plasma temperature of 1.2 $\times 10^{5}K$
(Table~\ref{tbl-4}). The 
measured Ly$\alpha$/RRC flux ratio for \ion{Ne}{10} was 1.9 in eclipse and 2.1
in $\phi$ 0.5, which is not consistent with the ratio of 1.3 predicted for a purely recombining
plasma, and implies photoexcitation is present \citep{shu02a}.

\subsection{The fluorescent lines.} 
There are 9 NS systems in the OB-supergiant category of HMXBs, to which Vela
X-1 belongs, and 2 
black hole systems \citep{kap01}. Of 5 systems for which high-resolution X-ray
spectra 
are available, multiple fluorescent lines are present in 3: Vela X-1 (this
study, 
\citet{shu02a}), GX 301-2 (our study in progress) and 4U 1700-37 \citep{bor03}.
Our study of GX 
301-2 reveals \ion{Ca}{2}-\ion{}{7} at 3.358 \AA, \ion{Ar}{6}-\ion{}{9} at 4.18 \AA,
\ion{Si}{9} at 5.316 \AA, \ion{S}{5} at 3.367 $\mbox{\AA}$ and \ion{Si}{5} at 7.116 \AA,
in addition to Ni and Fe
K$\alpha$ lines. Fe K$\alpha$ transitions are common in photoionized X-ray
spectra, but otherwise
a multitude of  fluorescent lines is unusual in X-ray astronomy.
\citet{mat97} suggest that fluorescent emission from elements other than iron
from neutral, externally X-ray illuminated matter may occur if the direct
radiation is blocked and
only reflected radiation is visible; then neutral ions of Ne, Mg, Si, S, Ar,
Ca, Cr and Ni may be
detected, and the line flux increases with increasing inclination angle of the
gas slab. Thus in 
both Vela X-1 and GX301-2 the Fe K$\alpha$ flux may vary in orbit as the angle
changes, and when 
it is maximal the Ni K $\alpha$ line (and other fluorescent lines) are most
apparent 
(Table~\ref{tbl-2}).

Photoionization and Compton ionization by high-energy X-ray or Gamma-ray
photons are effective ways
to remove inner shell electrons from an atom or ion, and the vacancy in the
inner shell is filled by
electron cascades from higher level shells of the ion.  The process involves
either a radiative 
transition - fluorescence - or a non-radiative transfer of energy to an
electron (Auger effect), 
so the fluorescence yield, defined as the probability that radiation is emitted
during the 
filling of the vacancy, is less than 1. The probability scales as $Z^4$, so
high-Z K$\alpha$ 
lines such as Fe are more common \citep{lie01}. The Ni K$\alpha$ emission line
in Vela X-1 was 
first detected by \citet{sak99}, and is present in the scattered component of
the eclipse and  $\phi$ 0.5 spectra. This line is rarely detected in X-ray binaries or X-ray
astronomy in general, 
and the first unambiguous detection in an active galactic nucleus was reported
in 2003
\citep{mol03}. 

A K$\alpha$ line occurs where an inner-shell ionization event causes a vacancy
in the 1s shell, 
and the electron making a transition to fill the vacancy is a 2p electron. By
contrast in 
collisional plasmas the processes of collisional excitation and
excitation-auto-ionization mostly 
affect the outer electrons, and are less often associated with inner-shell
auto-ionization 
\citep{kaa93}. 

Fluorescent lines in Vela X-1 are an indicator of cooler clumps in the wind.
\citet{sak99} 
developed a differential emission measure distribution to model Vela X-1, which
has 
well-constrained system parameters, and achieved a realistic reproduction of
the ASCA spectrum. 
The model had to accomodate lines from near-neutral Mg, Si, S, Ar, Ca, Fe, and
Ni. In a simple 
homogenous wind model the local ionization parameter in the emission region was
far too high to 
support near-neutral ions. The fluorescent lines must be produced in dense
clumps or filaments in 
the wind that may be 100 times denser than in the ambient wind, and have local
ionization 
parameters that are up to 2 orders of magnitude lower so they can support ions
of low
charge state. \citet{lie99} has described in irradiated accretion disks how
hard X-ray photons 
can "burrow" into cool material, and are absorbed through K-shell
photoionization producing 
fluorescent emission.

\subsection{Doppler Shifts of Emission lines}

The orbital velocity of the NS was calculated at 300 km $s^{-1}$, consistent
with \citet{sak99}. In $\phi$ 0.25, the lines seen in absorption (directly in
line of sight of the NS) generally had a blueshift consistent with this value
  (Tables~\ref{tbl-2},~\ref{tbl-3}). However other lines seen in emission (as a result
of scattering) had a redshift of the order of 200 km $s^{-1}$, indicating
their source in ions in part of the stellar wind with a relative bulk motion
of around 500 km $s^{-1}$.  

In eclipse the H- and He-like ions demonstrate a modest redshift, and 
in $\phi$ 0.5 they are blueshifted. The lines are produced in a process
(resonant scattering) whereby ions in the wind absorb a photon from the X-ray
source, and emit a photon in the line of sight to the observer. In eclipse
these ions are located in regions of the wind (from the supergiant) which are
receeding, and out-of eclipse are  approaching the observer. The mean
blueshift was 179 km $s^{-1}$ in 13 lines in $\phi$ 0.5 where fits with narrow
($<$ $\pm{200}$ km $s^{-1}$) confidence limits were obtained. This indicates
that the supergiant's wind velocity is less than expected, since at the
distance of the NS from the supergiant's surface, it should be around 
860 km s$^{-1}$, that is around 50\% of the wind terminal velocity of
1700 km $^{-1}$ \citep{dup80}). The result suggests that the high
photoionization conditions effectively quench the driving force and
acceleration of the wind, contributing to the formation of a Stromgren
zone (see section 4.5 below).

Precise line-shifts were difficult to obtain in the fluorescent ions, with the
exception of the Fe K$\alpha$
line. The Fe K$\alpha$ line does not follow the same pattern as H- and He-like
ions: although there is a small redshift apparent in eclipse, in $\phi$ 0.5
there is no blueshift as found in the H- and He-like ions.
This is consistent with the above argument that the fluorescent ions arise in
different regions of the stellar wind. The Fe K$\alpha$ line has a redshift
around 200 km $s^{-1}$ in $\phi$ 0 and 0.5, and a small blueshift at $\phi$
0.25. This shift is consistent with lightly ionized \ion{Fe}{8}-\ion{}{10} 
\citep{hou69}, and the orbital motion of 300 km $s^{-1}$ may explain the
blueshift at $\phi$ 0.25.

\subsection{Absorption and Emission lines at $\phi$ 0.25}

\citet{hab89} noted that when the X-ray source of HMXBs is viewed through the
stellar wind captured by the compact object, absorption K-edges such as O, Si,
S, and K- and L- edges of Fe are seen. Many of these edges are
apparent in $\phi$ 0.25, but only the K-edges of S and Fe are seen in
$\phi$ 0.5 (Table~\ref{tbl-5}).

Several lines seen in emission in eclipse appear in absorption at $\phi$ 0.25
(\ion{Ne}{10} Ly$\gamma$, the r lines of the Mg XI (Figure 2,
Table~\ref{tbl-3}) and Si XIII triplets (Table 3), and Mg V (Table 2). These
features only became clear when the smoothing technique was applied. Smoothing (or
binning) on the scale of the resolution make weak absorption and emission features
easily visible. The photons from the X-ray source in the line of sight may be scattered or
absorbed by the stellar wind en route to the observer. \citet{mil02} also noted these
triplets with r in absorption and i and f in emission in another supergiant HMXB, the
black hole system Cyg X-1.  We expect the phenomenon may also be found in other X-ray
binaries such as 4U 1700-37, when spectra at various orbital phases are available. Some
emission lines are seen at 0.25 as a result of resonant scattering
(\ion{Ne}{10} and \ion{Mg}{12} Ly$\alpha$ and Ly$\beta$ lines, \ion{Ne}{9}
triplet), although others may be obscured by the continuum. The Doppler shifts
of the lines seen in absorption and those seen in emission are different
reflecting their different regions of origin.

In $\phi$ 0.5 no absorption features are detected, apart from edges at 5.0 and
1.74 \AA (Figure 3). Phase 0.25 has numerous absorption lines and edges, so a further study
between these 2 phases may be warranted to find a transition phase with more
absorption features.

\subsection{Strong Emission Lines at $\phi$ 0.5 }

It is generally considered in HMXBs such as Cen X-3 and Vela X-1 that the
emission lines are most prominent in eclipse, but persist outside of the
eclipse phase of the orbit \citep{mil02}. We consider the marked absorption of
the continuum at $\phi$ 0.5 (Figure 7) to be responsible for the unexpectedly
strong emission lines seen in this phase. In eclipse only X-rays
which have been reprocessed in a visible part of the wind are observed. The
$\phi$ 0.5 line emission displays the same features as in $\phi$ 0, only fluxes
are stronger. Thus at $\phi$ 0.5 there is an "eclipse-like" condition with the
disappearance of the X-ray continuum in the band 4-12 $\mbox{\AA }$ associated with the
detection of the strong line emission due to scattering of photons in the
stellar wind.

SMC X-1 is a further example of stronger line emission apparent when the
continuum is blocked out of eclipse \citep{vrt01}. In SMC X-1 when
out-of-eclipse in the high state the X-ray source is not blocked and the
observer sees only a continuum (similar to the Vela X-1 $\phi$ 0.25 spectrum),
but when the source is blocked by precession of the disk (signified by the
low-state) the line emission is apparent \citep{vrt01}. Vela X-1
has no precessing disk, but the absorption of soft X-rays and other radiation
bands around $\phi$ 0.5 has been reported by many investigators.

\citet{lie01} note that if local material intervenes with our line of sight
to the continuum source, then absorption may produce observable photoelectric
edges or otherwise alter the shape of the continuum. The X-ray irradiated gas
in the vicinity of the continuum source radiates lines and continua
characteristic of the local conditions and may intercept continuum radiation
at energies corresponding to atomic level separations. Thus a photon can be
absorbed, and subsequently re-emitted at nearly the same energy, but with a
new direction, so a subtraction from the photon flux propagating toward the
observer may occur.

We postulate that a wind structure in the line of sight accounts for the
increased optical depth in the  3-12 $\mbox{\AA} $  band in $\phi$ 0.5. Two distinct
structures are described in the various reports of absorption (including those
that find a higher absorption after $\phi$ 0.5 than before it). These are the
photoionization wake, and the accretion wake (Figure 4). The X-ray source
creates a Stromgren zone in which the high photoionization conditions
effectively quench the driving force and acceleration of the wind. The
stagnant flow inside the ionization zone collides with the accelerating wind
at the trailing border to form a photoionization wake, as shown in
Figure 4. In the UV investigators have concluded the dip in the lightcurve
around $\phi$ 0.5 is due to the photoionization wake, because the source is
large (the supergiant) and the accretion wake could not sufficiently occlude
it. By contrast the X-ray source is very compact and is readily occluded by
the accretion wake at $\phi$ 0.5 (Figure 4). \citet{dum00} has reported how an
accretion wake in an X-ray nova has modified the lightcurve in the UV.

Supergiant stars have prodigious mass loss through stellar winds, typically
$10^{-6}$ $M_\odot$ per year. Wind material departs uniformly from the surface of
the star, and accretes onto the NS if it passes within a critical distance
forming an accretion column. Adjacent material that is perturbed in its flow
but is not accreted forms an accretion wake.

The discovery of this wind structure in Vela X-1 may be helpful in interpreting
emission lines in other X-ray binaries. 4U 1700-37 is also reported to have
strong emission lines out-of-eclipse ($\phi$ 0.7) \citep{bor03}. These authors
consider the possibility that the emission lines appear because the X-ray
source is seen "through an enhanced-density gas stream". However it may be
argued that the lines are seen via scattered photons. Analogous
to the case of Vela X-1, the lines may be seen in emission because the
continuum X-rays approaching directly from the source are absorbed by an
accretion stream or wake, allowing the scattered photons to be viewed. Haberl
et al. (1989) have modelled the accretion stream and wake in 4U 
1700-37, which was found to act as an absorber at $\phi$ 0.6.

\subsection{Source of X-rays}

\citet{rau94} has suggested that X-rays and gamma rays from Vela X-1 are 
produced by the same mechanism, that is by a beam of charged nucleons
accelerated in the strong magnetic field at some distance from the 
surface of the NS. The beam interacts with accretion material to form mesons 
which decay to gamma rays which  escape when the production region is far
enough from the NS so they are not depleted by pair production 
in the strong magnetic field, and the density is low enough to avoid 
absorption. However, \citet{lie01} and \citet{woj03} prefer resonant scattering of X-rays from an
accretion ``hot spot''. This study provides further support for resonant scattering.

Vela X-1 may have an accretion column, as well as an accretion wake. Vela 
X-1's X-ray signal is strongly pulsed, indicating there is a hot spot on the NS
surface. A hot spot is predicted from the cyclotron scattering resonance
features with a cyclotron resonance high-energy cutoff of 15-20 keV,
corresponding to a magnetic field strength of 2.3 $\times 10^{12}$ Gauss
(White et al., 1995). The hot spot is a thermal radiator and the pulse profile
is determined by the variation in the projected area of the hot spot as the 
NS rotates \citep{lyn98}. The pulse has high amplitude (amplitude 45\% in the 
1.2-2.3 keV band and 55\% in the 13.5-18.5 keV band of the time-averaged
signal), and a notch or "double pulse shape" in the main pulse in the X-ray
and UV pulse profiles (shown in Figure 7 of \citet{bor96}). Pulse profiles at 
different energies are further determined by shadowing of surface radiation by 
the accretion column and gravitational focusing of column radiation. A notch in
the main pulse may result where the accretion column is optically thick,
causing scattering of radiation propagating vertically through the accretion
column.

Insight as to how these complex structures interact to produce a strong 
absorption of the 5-12 $\mbox{\AA }$ band may be provided by study of pulse-phase resolved
emission. In a study of the HEG spectra \citet{pau02} demonstrated 
weak phase dependence of the Fe K$\alpha$ line and suggested that this 
represented an anisotropic velocity distribution; further studies on additional
lines are in progress.  

\section{Conclusions} 

X-ray emission in Vela X-1 results from capture and accretion of gas in the
stellar wind of the supergiant star onto the hot spot on the NS. Wind material
departs uniformly from the surface of the star, and accretes onto the NS if it
passes within a critical distance forming an accretion column. Adjacent
material that is perturbed in its flow but is not accreted forms an accretion
wake. The X-rays ionize and heat the surrounding gas, and are reprocessed in
the stellar wind. Spectra at orbital phases 0, 0.25 and 0.5 reflect the
processes of mass loss in the stellar wind of the supergiant and mass
accretion onto the NS. 

The Doppler shifts at $\phi$ 0.5 indicate slowing of the supergiant's wind
velocity in the vicinity of the NS, consistent with X-ray photoionization and
destruction of ions with strong UV resonance transitions that otherwise drive
the stellar outflow. The X-ray source creates a Stromgren zone in which the
high photoionization conditions effectively quench the driving force and
acceleration of the wind.   

Spectra at $\phi$ 0.25 reveal simultaneous evidence of emission lines produced
by scattering, and absorption lines that are produced in different regions of
the binary system. 

Narrow emission lines in Vela X-1 become apparent when the continuum is blocked
from line of sight, either by eclipse or by scattering and/or absorption from
a wake or cloud. The X-ray source is very compact and is readily occluded by
the accretion wake at $\phi$ 0.5, accounting for the absorption of soft
continuum. Absorption of the continuum at $\phi$ 0.5 has allowed detection of
scattered radiation and emission lines whose strength is an order of magnitude
higher than those in the eclipse phase. The eclipse and 0.5 spectra have a
similiar appearance, and the ratios of lines in the He-like triplets are
consistent with photoionization and resonant scattering (involving
photoexcitation) in a PI plasma. 

Multiple fluorescent emission lines have been detected in 3 of 5 supergiant HMXBs
for which high resolution X-ray spectra are available. This study suggests
fluorescent emission from externally X-ray illuminated matter may occur from elements other than
neutral iron if the direct radiation is blocked and only reflected radiation is
visible, when neutral ions of Ne, Mg, Si, S, Ar, Ca, Cr and Ni may be detected.

Differences in Doppler shifts of various lines indicate different origins of
lines, from direct radiation from the surface of the NS (lines seen in $\phi$
0.25 in absorption) and from adjacent regions in the stellar wind from
scattered radiation. 

\acknowledgments

DPH was supported by contract SAO SV1-61010 to MIT in support of the Chandra 
X-Ray Center.

We thank John Davis for assistance in the assessment for pileup in ObsId 1928. We
also wish to thank Chandra X-ray Observatory Center Director Harvey Tananbaum, and the
staff for executing these observations and their help in processing the data.
This research has made use of the data and resources obtained through the HEASARC on-line
service, provided by NASA-GSFC.

\clearpage

\begin{figure}
\plotone{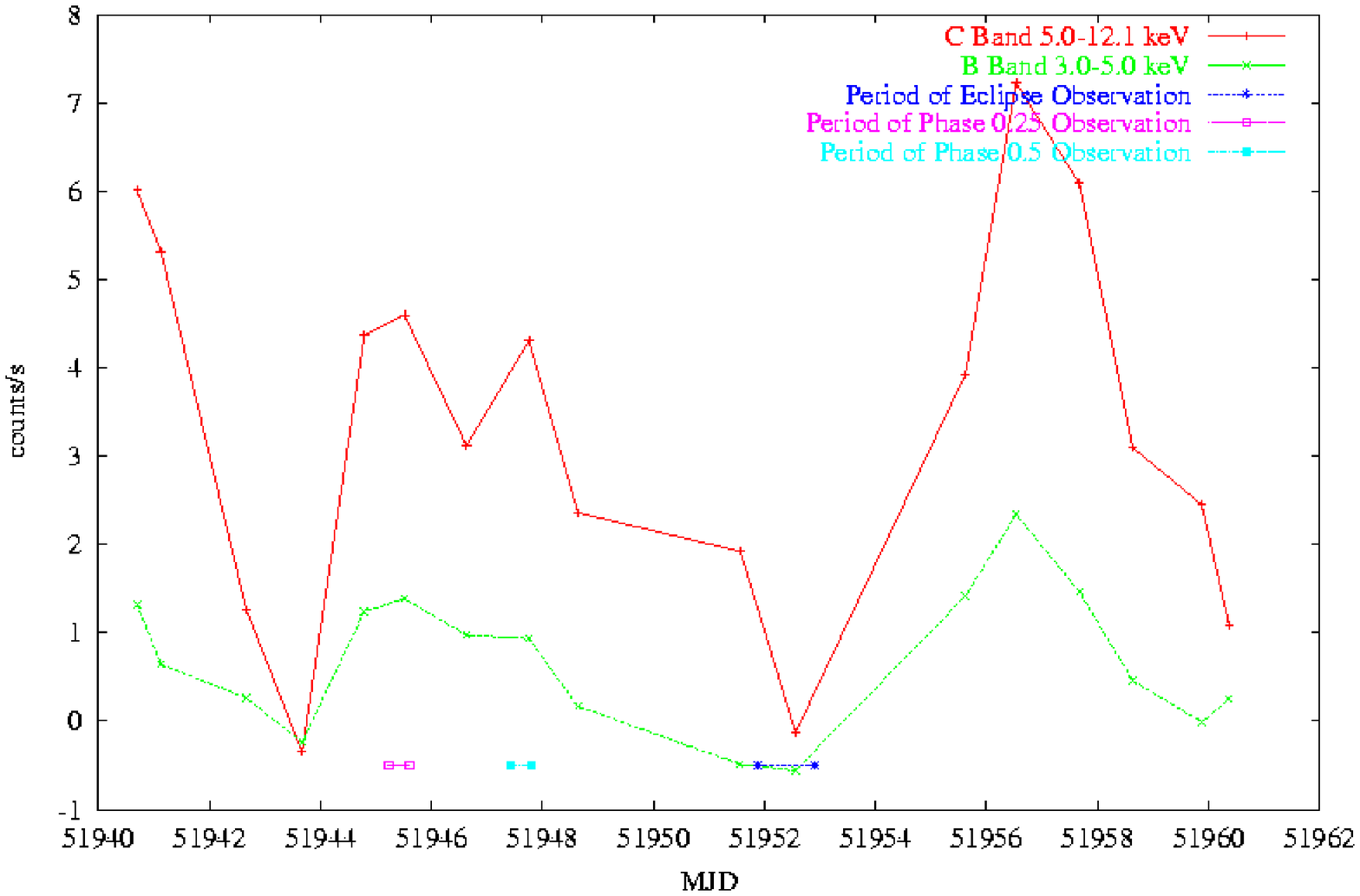}
\caption{RXTE ASM lightcurve of 2 orbits of the neutron star in Vela X-1,  showing the periods
(epochs) of the 3 Chandra datasets ( phases ($\phi$) 0, 0.25 and 0.5)
within one 8.96 day orbit. \newline \newline Fig. 2 NEXT PAGE
Comparison of smoothed spectra, showing the strong continuum in phase
($\phi$) 0.25 (upper spectrum) that partially obscures the emission
lines apparent in phase 0.5 (panels 2 and 3). There are 2 curves for
each phase, representing the high and medium energy grating (HEG and
MEG) spectra. In $\phi$ 0.25 HEG (black line) with higher resolution
has more sharp unresolved features than  MEG (red line), and its lower
area gives it larger statistical uncertainty. For $\phi$ 0.25 spectra,
the statistical uncertainties are about 6\% and 11\% for MEG and HEG,
respectively, near 7.5 \AA, and rise to about 18\% and 32\% near 11.5
\AA. There are three lines (\ion{Mg}{11} triplet r, \ion{Ne}{10}
Ly$\gamma$ and Ly$\beta$) in absorption in $\phi$ 0.25 and in emission
in $\phi$ 0.5. The inverted \ion{Mg}{11} triplet r is blueshifted
compared to the i and f lines.} \label{fig1} \end{figure}

\clearpage
\begin{figure}
\plotone{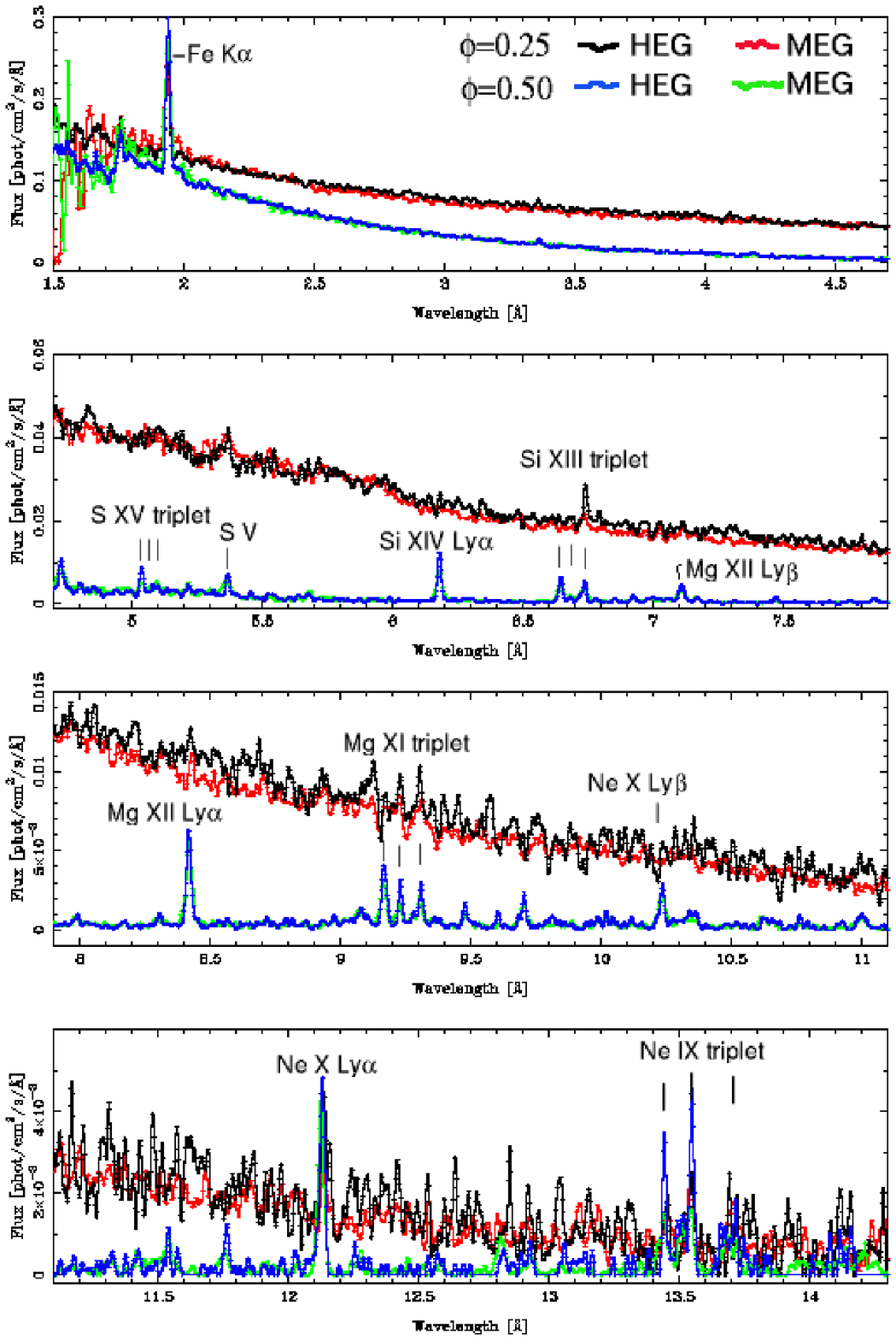}
\caption{Comparison of smoothed spectra}
\label{fig2} \end{figure}

\clearpage
\begin{figure}
\plotone{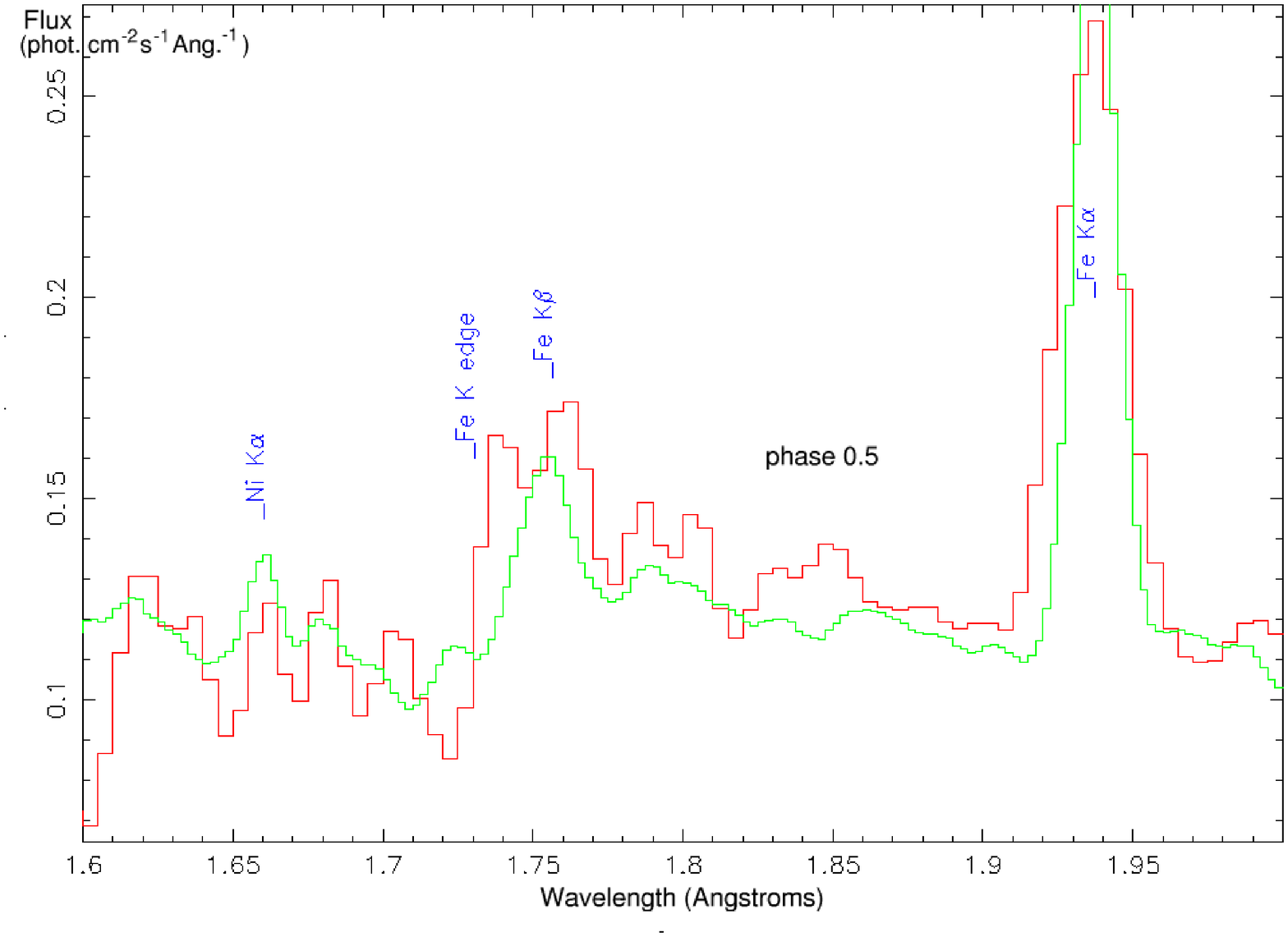}
\caption{$\phi$ 0.5 smoothed spectrum, showing Fe K$\alpha$, Fe K$\beta$ and Ni
K$\alpha$ lines, and Fe K edge. HEG is green and MEG is red.}
\label{fig3}
\end{figure}

\clearpage
\begin{figure}
\plotone{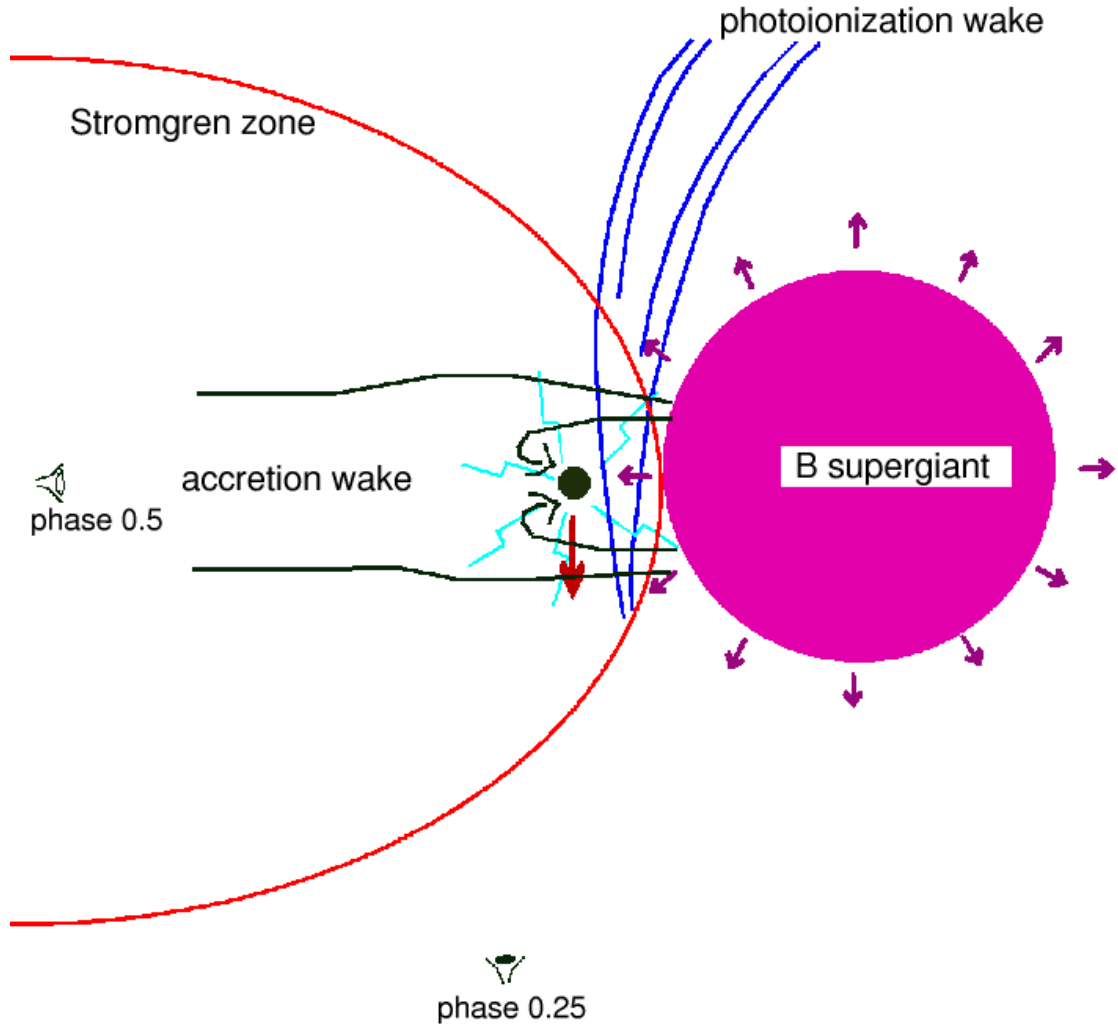}
\caption{Line of sight at orbital phases 0.25 and 0.5 of the neutron star and
its accretion wake and photoionization wake. The compact source on the neutron
star may be occluded by the accretion wake at $\phi$ 0.5. Wind material within a critical radius
accretes onto the neutron star (curved arrows), while material peturbed in its flow that does not
accrete forms an accretion wake.}
\label{fig4}
\end{figure}

\clearpage
\begin{figure}
\plotone{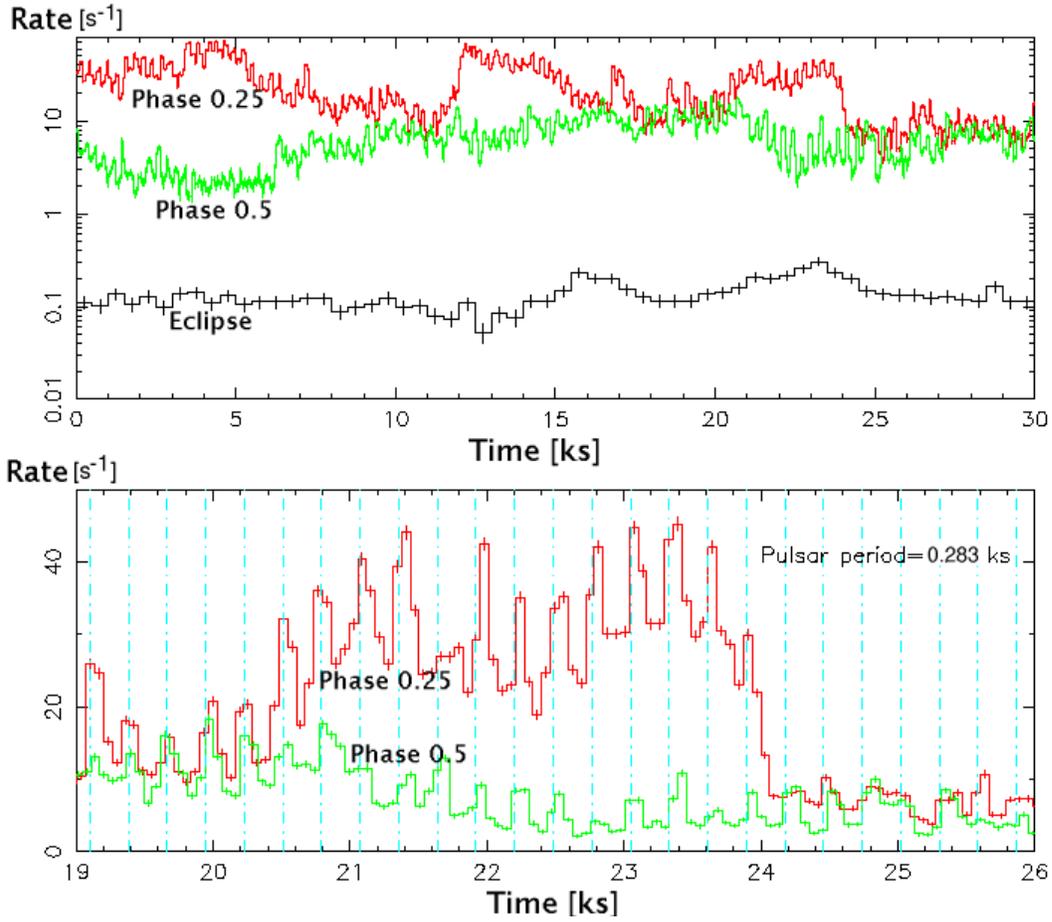}
\caption{X-ray lightcurves from the eclipse and $\phi$ 0.25 and $\phi$ 0.5
datasets in which pulsed emission (period 283 s) is apparent.}
\label{fig5}
\end{figure}

\clearpage
\begin{figure}
\plotone{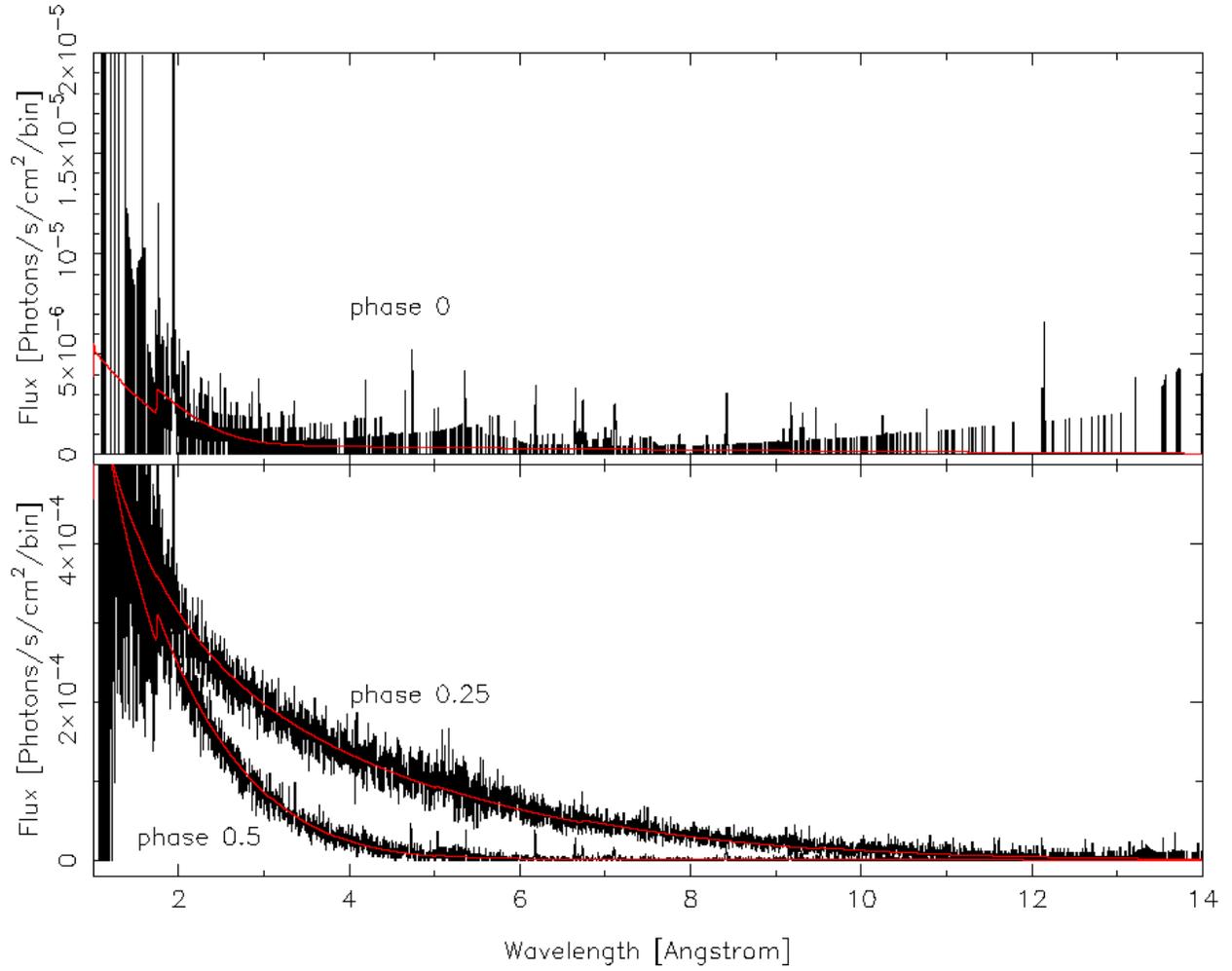}
\caption{Fitted spectra from 3 phases (eclipse, $\phi$ 0.25 and $\phi$ 0.5),
using a
best-fit model of 2 absorbed powerlaws.}
\label{fig6}
\end{figure}

\clearpage
\begin{figure}
\plotone{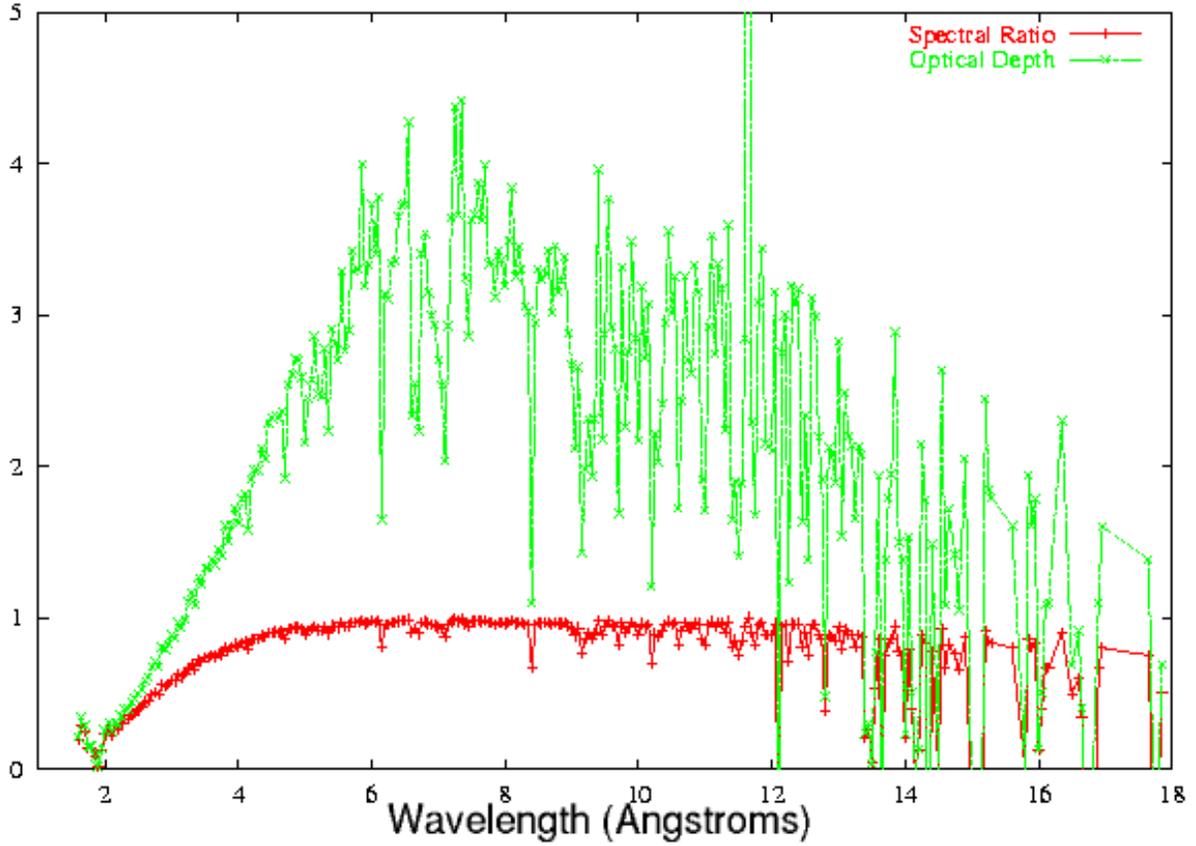}
\caption{Assessment of continuum absorption in $\phi$ 0.5 by comparison with
$\phi$ 0.25, using "spectral ratio" and optical depth models (see Observations
section for equations). The optical depth model (top curve) increases sharply from 2-6 \AA, then
gradually declines 6-13 \AA, then declines sharply, while the spectral ratio curve shows a gradual
increase from 2-4 \AA, and then is completely flat from 4-18 \AA.}
\label{fig7}
\end{figure}

\clearpage

\renewcommand{\arraystretch}{.8}

\begin{deluxetable}{lllll}
\tabletypesize{\scriptsize}
\tablecaption{Continuum Parameters in Eclipse, $\phi$ 0.25 and $\phi$ 0.5
\label{tbl-1}}
\tablewidth{0pt}
\tablehead{
\colhead{Parameter} & \colhead{Eclipse(ObsId 102)} & \colhead{Eclipse(ObsID
1926)}& \colhead{$\phi$ 0.25 }& \colhead{$\phi$ 0.5 }}
\startdata
Scattered & & & & \\
$N_H$\tablenotemark{a}  & 0.5  & 0.5  & 0.5 & 0.5 \\
 
Powerlaw norm\tablenotemark{b} & 0.00060 & 0.00064 & 0.0083 & 0.0066  \\
Powerlaw PI & 1.7  & 1.7 &1.7 &1.7 \\
   & & & & \\
Direct & & & & \\
$N_H$ & 39.9  & 39.0 & 1.2 & 12.2 \\

Powerlaw norm & 0.0081  & 0.0074 & 0.22 & 0.12 \\
Powerlaw PI & 1.7  & 1.7 & 0.92 & 0.60 \\
 & & & & \\
Flux(0.5-10 keV)\tablenotemark{c} & 1.17e-11 \tablenotemark{d}& 1.13e-11 & 
3.16e-9 & 2.04e-9 \\
 & & & & \\
\cutinhead{Dataset Details}
ObsID & 102 & 1926 & 1928 & 1927 \\
Start Date & 2000-04-13 & 2001-02-11 & 2001-02-05 & 2001-02-07 \\
Duration (ks) & 30 & 85 & 35 & 30  \\
$\phi$ & 0.98-0.02 & 0.97-0.08 & 0.23-0.27 & 0.48-0.52 \\

 \enddata

\tablenotetext{a}{in units of $10^{22} cm^{-2}$}

\tablenotetext{b}{in units of $10^{-5}$ photons $cm^{-2} s^{-1}$}
\tablenotetext{c}{in units of ergs $cm^{-2} s^{-1}$}
\tablenotetext{d}{\citet{shu02a} reported a measurement of 0.88e-11 ergs 
$cm^{-2} s^{-1}$ on this dataset }
\end{deluxetable}

\clearpage

\begin{deluxetable}{ll|ll|ll|ll|l}
\tabletypesize{\scriptsize}
\tablecaption{Fluorescent emission lines at 3 orbital phases in Vela X-1. Only 
a few flourescent lines are 
found in $\phi$ 0.25, including the \ion{Mg}{5} line which is blueshifted and 
seen in absorption. At the 
bottom of the table a comparison is made of the 2 fluorescent Fe K lines and 
the rare Ni K$\alpha$ line in 
Vela X-1 and GX 301-2. \label{tbl-2}}
\tablewidth{0pt}
\tablehead{
\colhead{Ion\tablenotemark{a}} & \colhead{Expected $\lambda$} & 
\colhead{Ob 1926}      &  & \colhead{Ob 1928} &  & \colhead{Ob 1927} &  & 
\colhead{Ratio}}
\startdata
  & \AA & $\phi$ 0 &   &  $\phi$ 0.25 &  & $\phi$ 0.5  & & Flux   \\
  &  & Shift\tablenotemark{b} & Flux\tablenotemark{c}  &  Shift & 
Flux\tablenotemark{d}  & Shift & Flux  & 
1927/26 \\
 &    &      &     &  &  &    &       &     \\

Ca II-VII & 3.359 &  -119 & 0.5 &\nodata  &\nodata  &$120_{-227}^{213}$  & 
$11.5_{5.3}^{5.3}$  & 22.6  \\
Ar VI-IX  & 4.186 & 847 & 0.2 &\nodata  &\nodata  &-82  &3.9  & 20.5  \\
\ion{S}{9} & 5.320 &  198 & 0.5 &\nodata          &\nodata & -377 & 2.2 & 4.2 
\\
\ion{S}{5} & 5.367 &  34  & 0.5 & $63_{-251}^{230}$& 11.6   & -150  & 3.8 & 
7.1 \\
Si XI-XII & 6.750-6.813   &\nodata  & 0.1 &\nodata  &\nodata  &\nodata  & 1.8 
& 18.0 \\
Si X-XI   & 6.813-6.882   &\nodata  & 0.3 &\nodata  &\nodata  &\nodata  & 0.9 
& 3.0 \\
\ion{Si}{9}& 6.947 & $-570_{-490}^{271}$ & $0.5_{-0.36}^{0.44}$ &\nodata  
&\nodata  & $-1028_{-137}^{275}$ & 
3.0 & 2.9 \\
\ion{Si}{8}& 7.007 & $ -119_{-488}^{389}$ & 0.3 &\nodata  &\nodata  & -396 & 
1.0 & 3.2 \\
\ion{Si}{7}& 7.063 &-85 & 0.2 &\nodata  &\nodata  & $-527_{-249}^{321}$ & 1.6 
& 8.2 \\
\ion{Si}{5}& 7.117 & \nodata  & \nodata &\nodata  &\nodata  
&$2076_{-233}^{149}$ & 3.2 &\nodata \\
Mg VIII-IX & 9.436-9.544&\nodata &\nodata & \nodata  &\nodata  &\nodata  
&$2.1_{1.0}^{1.0}$ &\nodata \\
\ion{Mg}{5} & 9.814 &-763 & 0.1 &$-578_{-264}^{303}$  &$-3.1_{-1.8}^{1.9}$  
&539  &0.6  & 5.4 \\
\ion{Mg}{2} & 9.890 & -217 & 0.5 &\nodata  &\nodata  & -10    &0.6  & 1.3\\

\cutinhead{Neutral Fe and Ni Fluorescent Lines in Vela X-1}
Ni K$\alpha$ & 1.660 & -328 & 4.860 & \nodata  &\nodata  &$103_{-374}^{369}$ 
&$74_{-33}^{52}$ & 15.1 \\
Fe K$\beta$  & 1.756 &$-47_{-582}^{905}$ & $2.2_{-1.5}^{1.6}$ 
&$41_{-315}^{384}$ & $59_{-28}^{28}$ 
&$-126_{-325}^{354}$ & $52_{-28}^{28}$ & 31.1 \\
Fe K$\alpha$ & 1.937 &$210_{-87}^{89}$ &$18.1_{-2.6}^{2.6}$ &$-67_{-68}^{67}$ 
&$221_{-19}^{19}$ & 
$119_{-48}^{48}$ & $374_{-27}^{26}$ & 20.7 \\
\cutinhead{Expected Line Shift Associated with Orbital Motion\tablenotemark{e}}
   &   &  -22  & \nodata & -322 & \nodata &  -22 & \nodata & \nodata\\
\cutinhead{Neutral Fe and Ni Fluorescent Lines in GX 301-2\tablenotemark{e}}
             &       & $\phi$ 0.1-0.2 &   &  $\phi$ 0.96-0.98    & & $\phi$ 0.47-0.48 &
& \\
Ni K$\alpha$ & 1.660 & \nodata  &\nodata  &$221_{-318}^{265}$ & 
$89_{-43}^{43}$ & 370  & 17 &\nodata \\
Fe K$\beta$  & 1.756 &$213_{-610}^{671}$ & $18_{-11}^{11}$ & 
$-58_{-109}^{109}$ &$220_{-24}^{24}$ & 
$-415_{-460}^{464}$ & $ 34_{-16}^{15} $ & \nodata \\
Fe K$\alpha$ & 1.937 &$258_{-90}^{91}$ & $88_{-9}^{8}$ & $285_{-25}^{25}$ 
&$788_{-24}^{22}$ &
$213_{-38}^{38}$  & $ 331_{-16}^{16}$ &\nodata \\

 \enddata

\tablenotetext{a}{from \citet{hou69}}
\tablenotetext{b}{in units of km $s^{-1}$; positive values are redshifted}
\tablenotetext{c}{in units of $10^{-5}$ photons $cm^{-2} s^{-1}$} 
\tablenotetext{d}{negative flux value indicates line is seen in absorption}
\tablenotetext{e}{for directly observed photons originating on or near the
surface of the neutron 
star}
\tablenotetext{f}{GX 301-2 results are from our study in progress}

\end{deluxetable}

\clearpage

\begin{deluxetable}{ll|ll|ll|ll|l}
\tabletypesize{\scriptsize}
\tablecaption{Flux and wavelength of H- and He-like lines at 3 phases 
\label{tbl-3}}
\tablewidth{0pt}
\tablehead{
\colhead{Ion} & \colhead{Expected $\lambda$\tablenotemark{a}} & 
\colhead{  Ob 1926}      &  & \colhead{  Ob 1928} &  & \colhead{  Ob 1927} &   
& \colhead{Ratio} 
}
\startdata
  &\AA  & $\phi$ 0 &  & $\phi$ 0.25 &  & $\phi$ 0.5  & & Flux   \\
  &  & Shift\tablenotemark{b} & Flux\tablenotemark{c}  &  Shift & 
Flux\tablenotemark{d}  & Shift & 
Flux  & 1927/26    \\
\cutinhead{H-like Ions}

\sidehead{\ion{S}{16}}
Ly$\alpha$& 4.730 & $397_{-389}^{342}$ & $1.2_{-0.6}^{0.6}$ &\nodata  &\nodata 
 &$-123_{-191}^{194}$ & 
$11.1_{-4.6}^{4.5}$ & 9.5 \\

\sidehead{\ion{Si}{14}}
Ly$\beta$& 5.217   & 322 & 0.3 &\nodata  &\nodata  & $24_{-297}^{275}$ & 
$4.9_{3.4}^{4.1}$ & 16.5 \\
Ly$\alpha$ & 6.183 & $334_{-142}^{145}$ & $1.2_{-0.66}^{6.74}$ &\nodata  
&\nodata & $-163_{-42}^{41}$ & 
$18.6_{-3.0}^{3.1}$ & 16.0 \\

\sidehead{\ion{Mg}{12}}

 Ly$\beta$ & 7.106 &  $333_{-60}^{59}$ & $1.2_{-0.32}^{0.33}$ & 
$311_{-131}^{101}$  &$6.2_{-3.5}^{3.7}$  & 
$44_{-64}^{45}$ & $8.8_{-2.0}^{2.7}$  & 7.1 \\

 Ly$\alpha$& 8.422 & $310_{-59}^{60}$ & $.96_{-0.51}^{0.67}$ 
&$154_{-176}^{165}$  &$5.3_{-2.0}^{4.6}$  
&$-153_{-30}^{43}$  & $13.9_{-2.0}^{2.0}$  & 14.4 \\

\sidehead{\ion{Ne}{10}}
 Ly$\delta$& 9.481 &  262 & 0.14 & -356  & -1.6  &$ -132_{-186}^{181}$ & 
$1.8_{-0.7}^{1.1}$ & 12.9 \\
 Ly$\gamma$  & 9.708 & 5 & 0.2 &$-39_{-360}^{215} $ &$-4.3_{-2.6}^{2.6}$  & 
$-225_{-123}^{120}$ & 
$4.6_{-1.6}^{2.2}$ & 24.2 \\
 Ly$\beta$ & 10.239& $ 337_{-323}^{327}$ & $0.4_{-0.36}^{0.37}$ & -186  & -1.7 
 &  $-190_{-86}^{87}$ & 
$6.3_{-1.7}^{5.0}$ & 15.6 \\
 Ly$\alpha$& 12.135 & $269_{-47}^{74}$ & $1.2_{-0.7}^{0.7}$ &$5_{-50}^{104}$  
&$3.9_{-2.3}^{2.4}$  
&$-264_{-55}^{52}$ & $13.7_{-2.7}^{2.7}$   & 11.9  \\
\sidehead{\ion{O}{8}}
 Ly$\alpha$ &18.970 &  563 & 0.48 &\nodata  &\nodata  & 73 & 1.6   & 3.4 \\
&    &      &     &  &  &    &       &     \\

\cutinhead{He-like Triplets}
\sidehead{\ion{S}{15}}
 r     & 5.039 &  401 & 0.97 &\nodata  &\nodata  & $-161_{-108}^{168}$ & 
$8.9_{-2.8}^{2.8}$ & 9.9 \\
 i     & 5.065 &  426 & 0.20 &\nodata  &\nodata  & -96 & 0.6 & 3.0  \\
 f     & 5.102 &  264 & $0.44_{0.4}^{0.4}$ &\nodata  &\nodata  & 
$-264_{-2174}^{1645}$ & 4.8 & 10.9 \\

\sidehead{\ion{Si}{13}}
 r    & 6.648 & $ 307_{-205}^{124}$ & $1.51_{-0.4}^{0.4}$ &$-477_{-293}^{275}$ 
 &$-3.9_{-0.9}^{2.7}$  & 
$-184_{-124}^{41}$ & $9.4_{-1.3}^{1.3}$ & 10.1\\
 i    & 6.687 &  287 & 0.1 & 152  &$6.0_{-3.6}^{3.6}$  & $-188_{-294}^{116}$ & 
$3.9_{-2.2}^{3.2}$ & 39  \\
 f    & 6.740 &  $227_{-293}^{218}$ & $0.93_{-0.37}^{0.36}$ 
&$-258_{-465}^{152}$  &$9.0_{-3.9}^{4.0}$  & 
$-242_{-123}^{57}$  &$9.4_{-2.3}^{1.0} $ & 10.6 \\

\sidehead{\ion{Mg}{11}}
 r    & 9.169 &  $343_{-152}^{120}$ & $0.89_{-0.8}^{0.79}$ 
&$-666_{-162}^{141}$  &$-5.2_{-1.2}^{3.6}$  & 
$-173_{-86}^{46}$ & $6.0_{-1.1}^{1.2}$ & 17.3 \\
 i    & 9.230 &  309 & 0.22 & -97  & $4.2_{-2.9}^{1.0}$  & -172   & 
$1.9_{-0.9}^{0.9}$ & 8.7  \\
 f    & 9.314 &  $354_{-553}^{341}$ & $0.35_{-0.3}^{0.3}$ &$-93_{-207}^{177}$  
&$4.1_{-2.3}^{2.3}$  & 
$-267_{-45}^{76}$ & $6.7_{-2.2}^{1.6}$  & 19.6 \\

\sidehead{\ion{Ne}{9}}
 r    & 13.447  &  201 & 0.34 & 359  & $1.1_{-0.5}^{1.3}$ & 
$-218_{-830}^{1400}$ & $1.1_{-0.7}^{0.9}$ & 2.6 
\\
 i    & 13.552  &  299 & 0.10 & $-153_{-208}^{219}$  & $4.0_{-2.4}^{2.4}$ & 
-199 & $1.6_{-1.0}^{1.3}$ & 16.1 
 \\
 f    & 13.699  &  328 & 0.42 & 241 & $2.1_{-0.9}^{2.7}$ & $-241_{-613}^{263}$ 
& $1.7_{-1.4}^{0.7}$ & 5.2 \\

 \enddata

\renewcommand{\arraystretch}{.6}
\tablenotetext{a}{from Chandra X-Ray Center Atomic 
Database(http://asc.harvard.edu/atomdb/)}
\tablenotetext{b}{in units of km $s^{-1}$; positive values are redshifted}
\tablenotetext{c}{in units of $10^{-5}$ photons $cm^{-2} s^{-1}$}
\tablenotetext{d}{negative flux values indicate lines that are seen in 
absorption}

\tablecomments{for H-like (and He-like i) doublets the mean $\lambda$ of the 2 
lines is used.}

\end{deluxetable}

\clearpage
\begin{deluxetable}{l|ll|ll}
\tabletypesize{\scriptsize}
\tablecaption{RRC Fit Parameters (redge\tablenotemark{a} ~model)\label{tbl-4} }
\tablewidth{0pt}
\tablehead{
\colhead{Parameter} & \colhead{Eclipse}& & \colhead{$\phi$ 0.5}&}
\startdata
    & \ion{Ne}{10}  &\ion{Ne}{9} & \ion{Ne}{10}  & \ion{Ne}{9}  \\
norm\tablenotemark{b}  & 0.6 & 0.6 & 6.5 & 6.0   \\
 
edge\tablenotemark{c} & 1.3628& 1.1956 & 1.3628 & 1.1956  \\
kT\tablenotemark{d} & 0.01035  & 0.01035  & 0.01035  & 0.01035  \\
 \enddata
\tablenotetext{a}{For \ion{Ne}{10} the complete fit function used was  
"gauss(1)+gauss(2)+gauss(3)+redge(1)+poly(1)", 
and for \ion{Ne}{9} "gauss(1)+redge(1)+poly(1)"; only redge parameters are 
displayed} 
\tablenotetext{b}{in units of $10^{-5}$ photons $cm^{-2} s^{-1}$}
\tablenotetext{c}{in units of keV}
\tablenotetext{d}{in units of keV  (kT of 0.01035 keV represents a temperature 
of 1.2 $\times$ $10^{5} K$)}
\tablecomments{The measured Ly$\alpha$/RRC flux ratio for \ion{Ne}{10} was 1.9 
in eclipse and 2.1 in $\phi$ 0.5}
\end{deluxetable}

\clearpage

\begin{deluxetable}{lrrr}
\tabletypesize{\scriptsize}
\tablecaption{Photo-electric Absorption Edges in $\phi$ 0.25 and
$\phi$ 0.5\label{tbl-5}}
\tablewidth{0pt}
\tablehead{
\colhead{$\phi$} & \colhead{Edge} & \colhead{$\lambda$(\AA)}   &
\colhead{Opt. Depth $\tau$}}

\startdata
$\phi$ 0.25   & S K    &5.00 &0.026  \\
             & Si K   &6.72 &0.054 \\
             & Mg K   &9.50 &0.097 \\
             & Ne K   & 14.30 &0.42  \\
             & Fe L3\tablenotemark{a}  & 17.55   &0.12   \\ 
$\phi$ 0.5    & S K    & 5.0  & 0.17   \\
             & Fe K\tablenotemark{b}   & 1.74 & 0.15     \\         
 \enddata

\tablenotetext{a}{Fe L3 edge is due to transition from the $2p_{3/2}$ level}

\tablenotetext{b}{Fe K edge is present at 1.74 $\mbox{\AA}$ in $\phi$ 0.5, and absent from
other phases, consistent with variation in the angle of the slab of gas with orbital phase.  This
edge is also present (and stronger $\tau$ = 0.90) in GX 301-2}
\end{deluxetable}

\end{document}